\documentclass[showpacs,aps,amsmath,amssymb,reprint,pre]{revtex4-1}

\usepackage{amsmath}
\usepackage{amssymb}
\usepackage{epsfig}
\usepackage{epstopdf}
\usepackage{amsthm}
\usepackage{natbib}

\usepackage{graphicx}
\usepackage{dcolumn}
\usepackage{bm}

\newcommand{\ave}[1]{\ensuremath{\langle {#1} \rangle}}

\begin{document}
	
	\title{Bursting endemic bubbles in an adaptive network}
	
	\author{N. Sherborne}
	\author{K. B. Blyuss}%
	\email{K.Blyuss@sussex.ac.uk}
	
	\author{I. Z. Kiss}
	\affiliation{Department of Mathematics, University of Sussex, Brighton, UK, BN1 9QH}
	
	\date{\today}
	
	\begin{abstract}
		The spread of an infectious disease is known to change people's behavior, which in turn affects the spread of disease. Adaptive network models that account for both epidemic and behavior change have found oscillations, but in an extremely narrow region of the parameter space, which contrasts with intuition and available data. In this paper we propose a simple SIS epidemic model on an adaptive network with time-delayed rewiring, and show that oscillatory solutions are now present in a wide region of the parameter space. Altering the transmission or rewiring rates reveals the presence of an endemic bubble - an enclosed region of the parameter space where oscillations are observed.
		
	\end{abstract}
	
	\pacs{89.75.Hc, 87.19.X-, 02.50.Ey}
	
	\maketitle
	
	The spread of an infectious disease changes the behavior of individuals, and this, in turn, affects the spread of the disease \cite{funk2009spread}. Broadly speaking, responses to an epidemic fall into two categories: coordinated and uncoordinated. Coordinated responses include vaccination and quarantine schemes, travel restrictions, and information spread through mass media. Uncoordinated responses cover individuals adapting their behavior based on their own perceived risk, this includes improved hygiene regimens and avoiding crowded places and public transport during outbreaks. Surveys consistently identify such precautionary measures taken by individuals during epidemic outbreaks \cite{rubin2009public, goodwin2009initial}. Fear of becoming infected during the 2003 SARS epidemic in Hong Kong caused huge behavioral shifts; air travel into Hong Kong dropped by as much as $80\%$ \cite{ferguson2006strategies}. Responses to a large study covering numerous European and Asian regions revealed that, in the event of an influenza pandemic, $75\%$ of people would avoid public transport, and $20-30\%$ would try to stay indoors \cite{sadique2007precautionary}. These behavioral shifts change the potential routes for transmission and can alter the size and time-scale of an epidemic \cite{funk2010modelling}. 
	
	In the context of epidemic models on networks, perhaps, the most widespread approach to couple epidemics and behavior is by using adaptive networks, where behavioral changes are captured by link rewiring based on the disease status of nodes \cite{gross2008adaptive, funk2010modelling}. Gross et al. \cite{gross2006epidemic, gross2008robust} considered a simple SIS model with rewiring, in which susceptible nodes disconnect from infected neighbours at rate $\omega$, and immediately reconnect to a randomly chosen susceptible node. This simple model led to bistability and to oscillatory solutions, albeit with oscillations limited to an extremely narrow region of the parameter space. This rewiring procedure has since been extended to consider scenarios where both the susceptible and infected nodes can rewire, and diseases with a latent period \cite{risau2009contact}. Zhang et al. \cite{zhang2012modeling} presented a further alternative, where news about past prevalence influences whether nodes choose to disconnect edges. The authors found an estimate of the critical delay that induces a Hopf bifurcation, thus causing periodicity. Tunc et al. \cite{tunc2013epidemics} studied a network model with temporary deactivation of edges between susceptible and infected individuals. On a growing network, Zhou et al. \cite{zhou2012epidemic} showed that cutting links between susceptible and infected individuals can lead to epidemic re-emergence, with long periods of low disease prevalence punctuated by large outbreaks. 
	
	Periodic cycles and disease re-emergence are evident in real-world data. Many diseases are subject to seasonal peaks, which have been studied extensively \cite{altizer2006seasonality,grassly2006seasonal}. Often a sinusoidal or other form of time-varying transmission parameter is used to imitate seasonality, which can lead to multiennial peaks \cite{kuznetsov1994bifurcation}. A number of models have identified other possible causes of periodicity in epidemic dynamics. To give one example, Hethcote at al. \cite{hethcote1981nonlinear} showed that in a well-mixed population temporary immunity in SIRS- or SEIRS-type models as represented by a time delay can result in the emergence of periodic solutions when the immunity period exceeds some critical value.
	
	One should note that seasonality alone cannot explain all cases of oscillations. In both the UK and the USA, the 2009 H1N1 pandemic occurred in two distinct waves separated by a few months \cite{campbell2011hospitalization,jhung2011epidemiology}. Other diseases have shown more long-term trends. Incidence reports of mycoplasma pneumonia have found evidence of epidemic cycles in many different countries, with periodicity of 3 to 5 years \cite{ito2001culturally,waites2004mycoplasma}. Recently, it has been suggested that syphilis exhibits periodic cycling \cite{grassly2005host}, although these findings have been subsequently questioned \cite{breban2008there}. Whilst it is difficult to pinpoint the specific causes of periodicity in the dynamics of these diseases, if syphilis epidemics are indeed cyclical, then changes in human behavior have been proposed as the likely explanation \cite{althouse2014epidemic}. 
	
	Intuitively, and as shown by empirical observations, one would expect oscillations to appear in epidemic models where behavior is considered. If an individual is aware of the state of their neighbours and responds accordingly, then times of high prevalence will be associated with greater caution, curbing further spread. Conversely, without advance warning, behavior will return to normal as prevalence wanes, enabling a second wave of the epidemic. Despite this intuition, adaptive network models have so far not been able to show such robust oscillations over reasonable regions of the parameter space. To tackle this problem, we introduce a simple SIS model on an adaptive network with $N$ nodes. Infected nodes transmit the disease to susceptible neighbours at rate $\beta$ across links, and recover and become susceptible again at rate $\gamma$, independently of the network. Susceptible nodes cut links that connect them to infected neighbours at rate $\omega$ and, after a fixed time delay of length $\tau$, reconnect to susceptible nodes chosen uniformly at random from all such available nodes. The delay between cutting and reconnecting is crucial. It is unrealistic to expect that alternative contacts can be identified and established arbitrarily quickly. The delay represents both people's hesitance to make new contacts and also the potential lack of availability of such new contacts when an epidemic is spreading thorough a population \cite{sadique2007precautionary}.
	
	To construct the mean-field model, we use the pairwise approximation method \cite{keeling1997correlation}. The number of nodes in the susceptible or infected state at time $t$ is denoted by $[S]$ and $[I]$, respectively; $[SS]$, $[SI]$ and $[II]$ denote the number of connected pairs of nodes in the respective states, with all pairs being doubly counted. The explicit dependence on time is dropped for simplicity. For the moment closure approximation we use the assumption that once a node is fixed, typically a susceptible node, then the 
	states of the neighbours are Poisson-distributed~\cite{rand1999}. This leads to:
	\begin{equation}\label{clos}
	[ABC] = \frac{[AB][BC]}{[B]},\quad A,B,C \in \{S, I\},
	\end{equation}
	to express the number of connected triples \cite{keeling1997correlation,gross2006epidemic}.
	
	The delay before an $S-I$ edge is rewired to an $S-S$ edge introduces a complication, as not all newly formed edges will be between two susceptible nodes. To see this, consider an example of a susceptible node with two or more infected neighbours. At some time $t_1$ it disconnects from one of these neighbours. Then, in the interval $(t_1, t_1 + \tau)$ another infected neighbour transmits the disease to it. If it then remains infected until time $t_1 + \tau$, the new edge will be of an $I-S$ type rather than $S-S$. To deal with this issue we use a technique similar to that used by Kiss et al.~\cite{kiss2015generalization} for a pairwise model with an infectious period of fixed length. Consider $y_p(t)$ to be the cohort of susceptible nodes that have cut a link at time $t - \tau$ and are waiting to reconnect. The expected number of infected neighbours a susceptible node has is approximated by $[SI]/[S]$. Therefore, the rate at which nodes in the cohort become infected over the interval $(t - \tau, t)$ is
	\begin{equation*}
	\dot{y}_p = -\beta y_p \frac{[SI]}{[S]}.
	\end{equation*}
	The solution to this ODE is
	\begin{equation}
	y_p(t) = \omega[SI](t - \tau)\exp\left( -\beta \int\displaylimits_{t-\tau}^t \frac{[SI](u)}{[S](u)} \, du \right),
	\label{eq:yp}
	\end{equation}
	since $y_p(t - \tau) = \omega[SI](t - \tau)$. 
	
	\begin{figure}
		\includegraphics[width = \columnwidth]{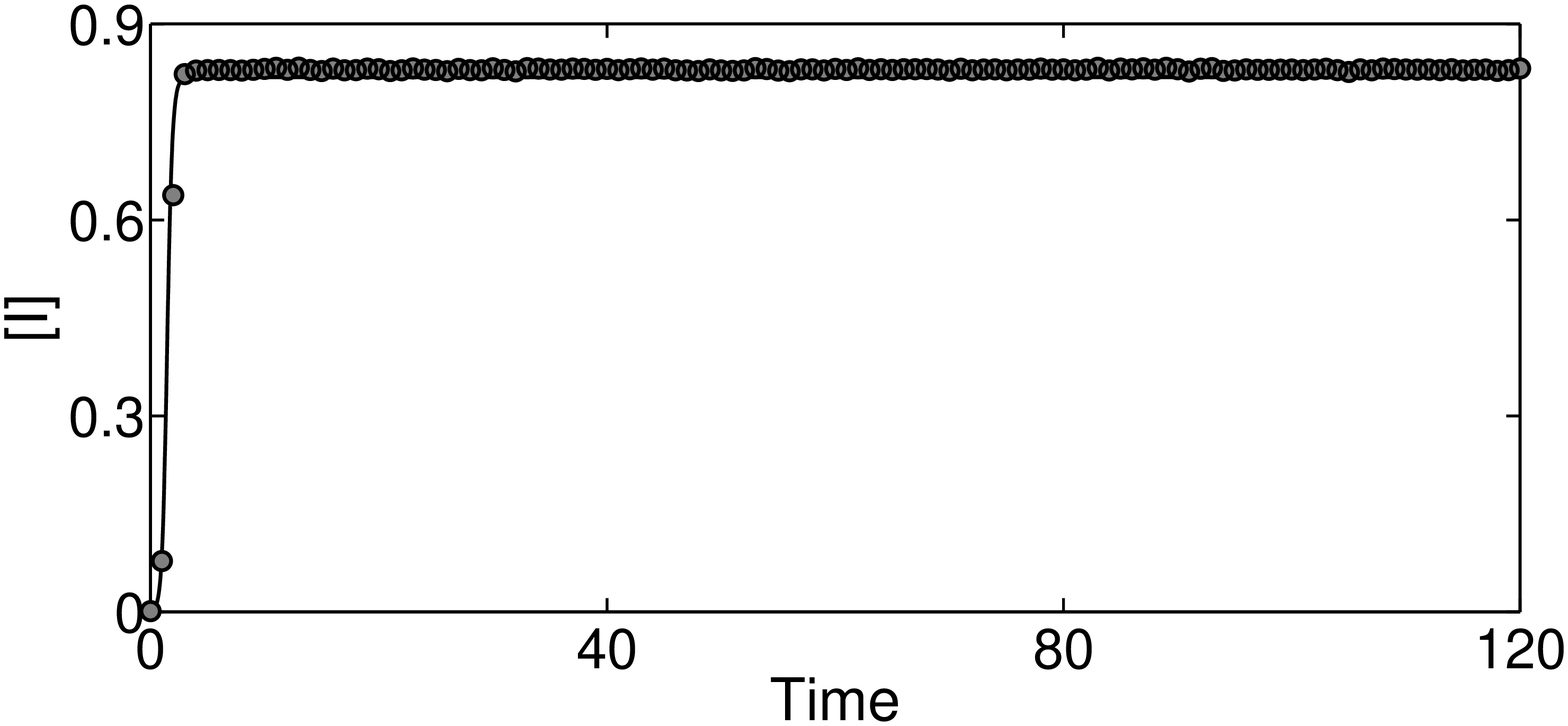} \\
		\includegraphics[width = \columnwidth]{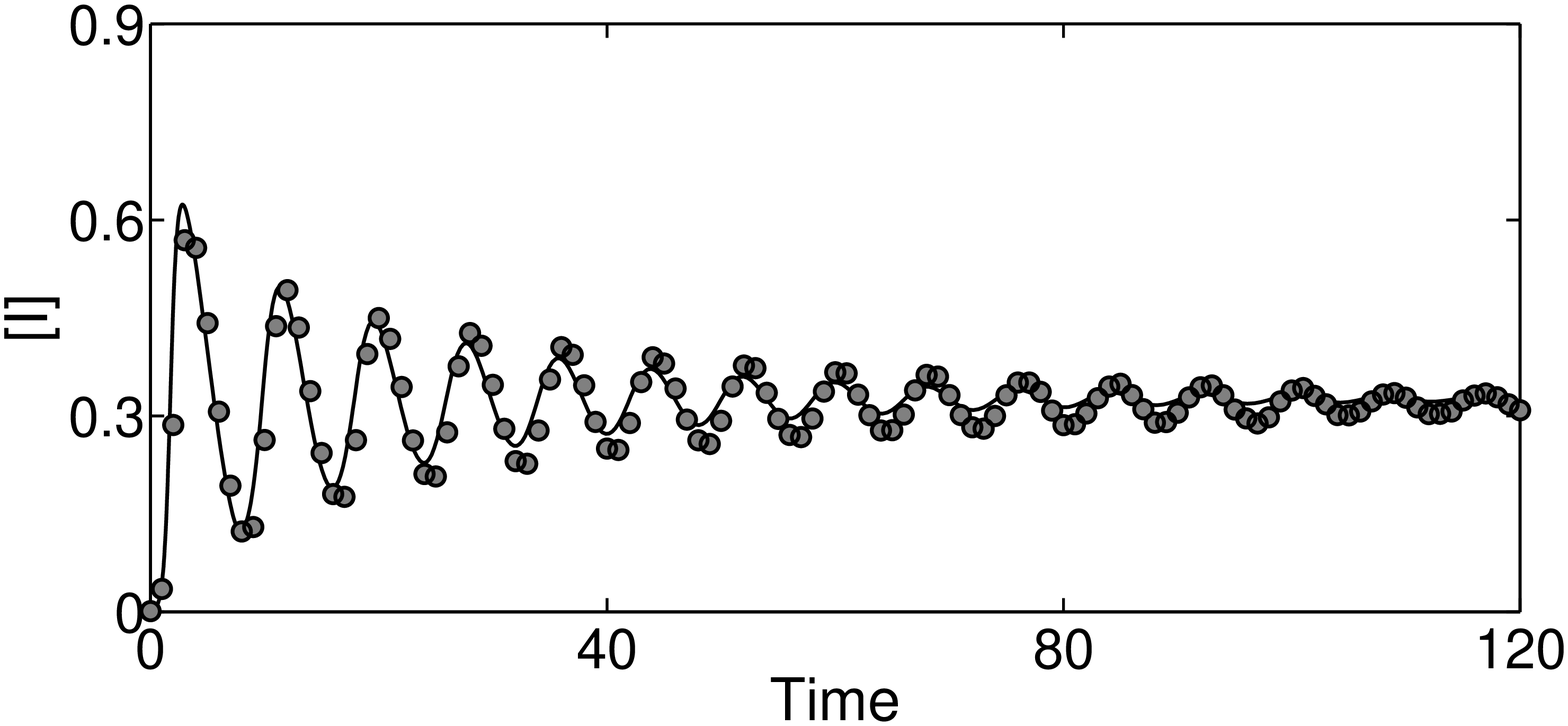} \\
		\includegraphics[width = \columnwidth]{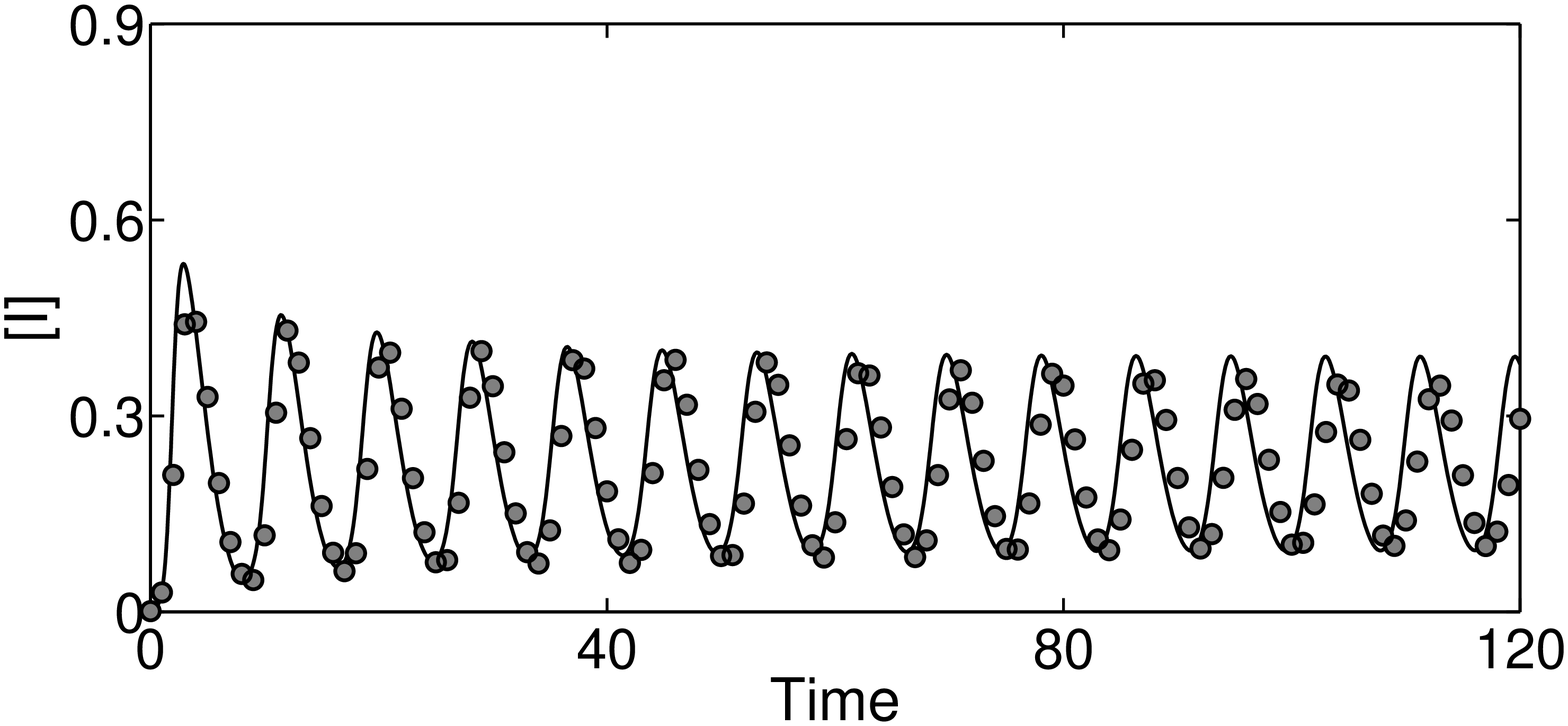} \\
		\caption{Comparison between the solution of~\eqref{eq:sisRewire} and numerical simulation. Three sets of results are shown, $\omega = 0$ (top), $\omega = 1$ (middle) and $\omega = 1.4$ (bottom). Other parameters are $\beta = 0.6$, $\gamma = 1$, $\tau = 6$ and $\ave{k} = 10$. Simulation results are averaged across 100 iterations on random networks of 1000 nodes. All simulations begin by randomly selecting a node to infect at time $t=0$. Simulation runs which die out are discarded and performed again.}
		\label{fig:numSim}
	\end{figure}
	A member of the cohort infected at some time $u \in (t - \tau, t)$ may recover before time $t$. To ensure that we only consider nodes which remain infected, we must include the probability that a node infected at time $u$ remains infected until time $t$ in the integral term of \eqref{eq:yp}. This is the survival probability of the recovery process, and it is given by $e^{-\gamma(t - u)}$. Therefore, the rate at which new $S-S$ edges are formed is,
	\begin{equation}
	y(t) := \omega[SI](t - \tau)\exp\left( -\beta \int\displaylimits_{t-\tau}^t \frac{[SI]}{[S]}e^{-\gamma(t - u)} \, du \right).
	\label{eq:Discount}
	\end{equation}
	\begin{figure*}[htp]
		\centering
		\includegraphics[trim = 20 0 175 0, clip,width=.47\textwidth]{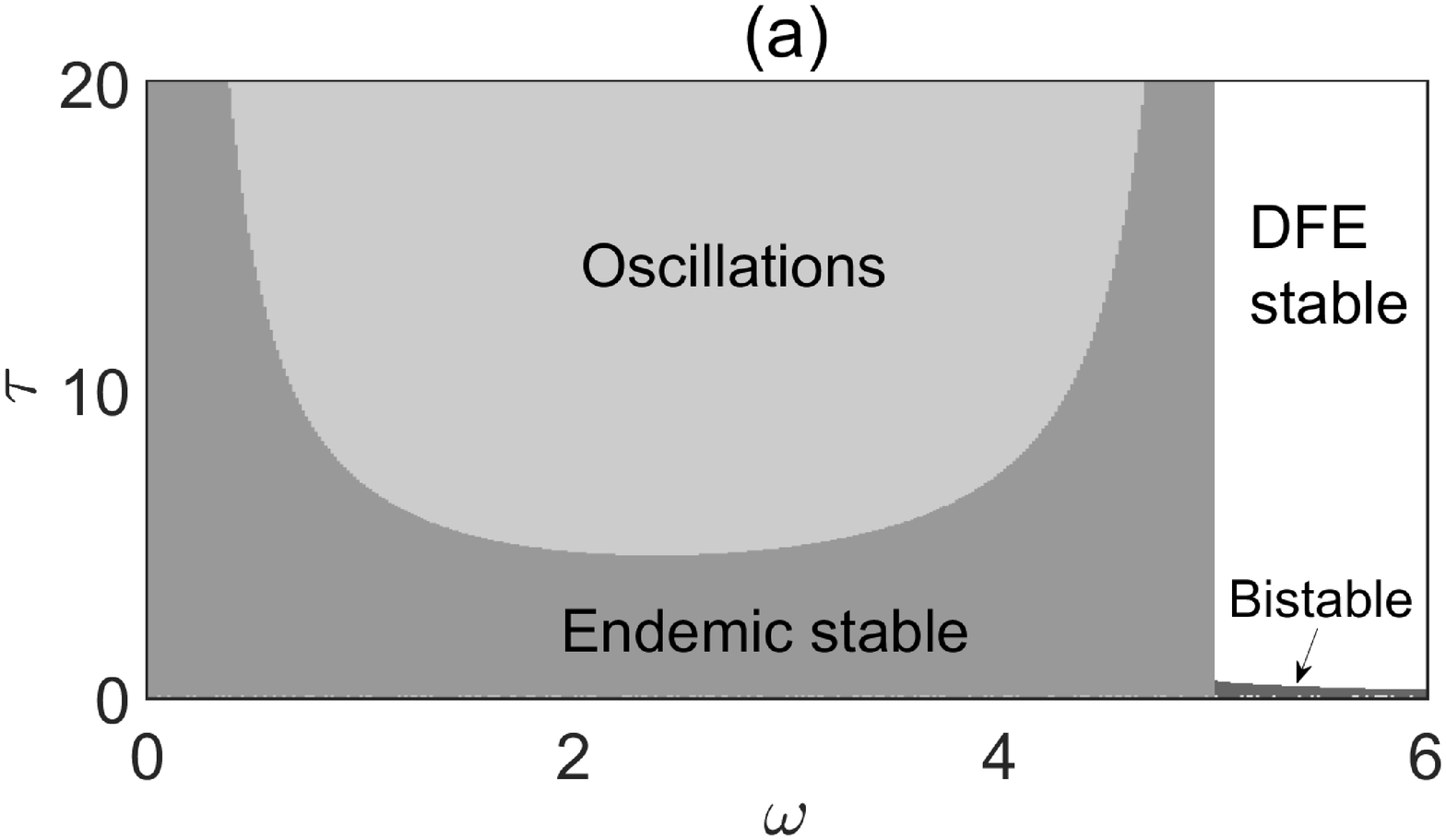}\quad
		\includegraphics[trim = 20 0 135 0, clip,width=.49\textwidth]{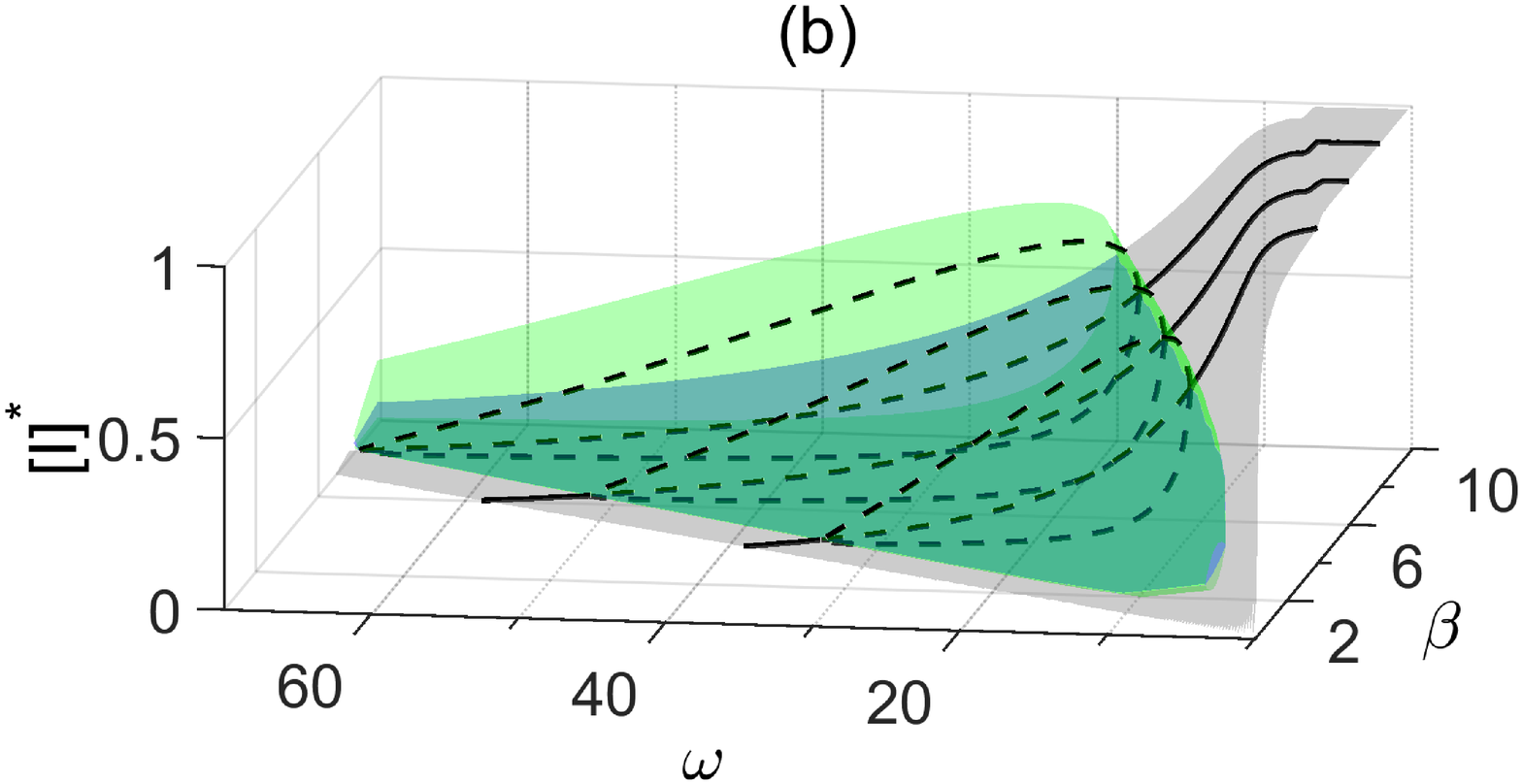}\quad
		
		\medskip
		\includegraphics[trim = 35 0 130 0, clip, width=.47\textwidth]{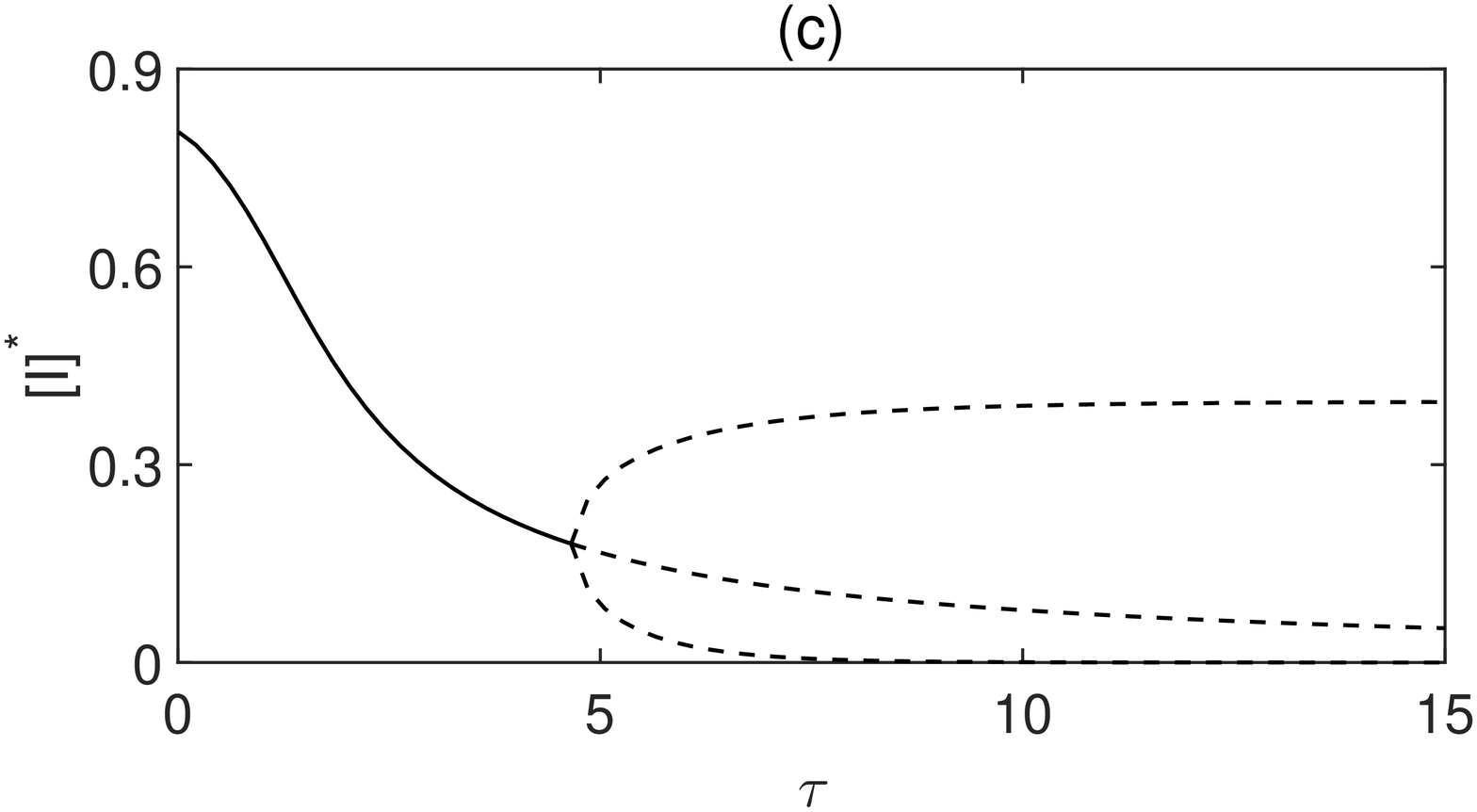}\quad
		\includegraphics[ trim = 35 0 130 0, clip, width=.47\textwidth]{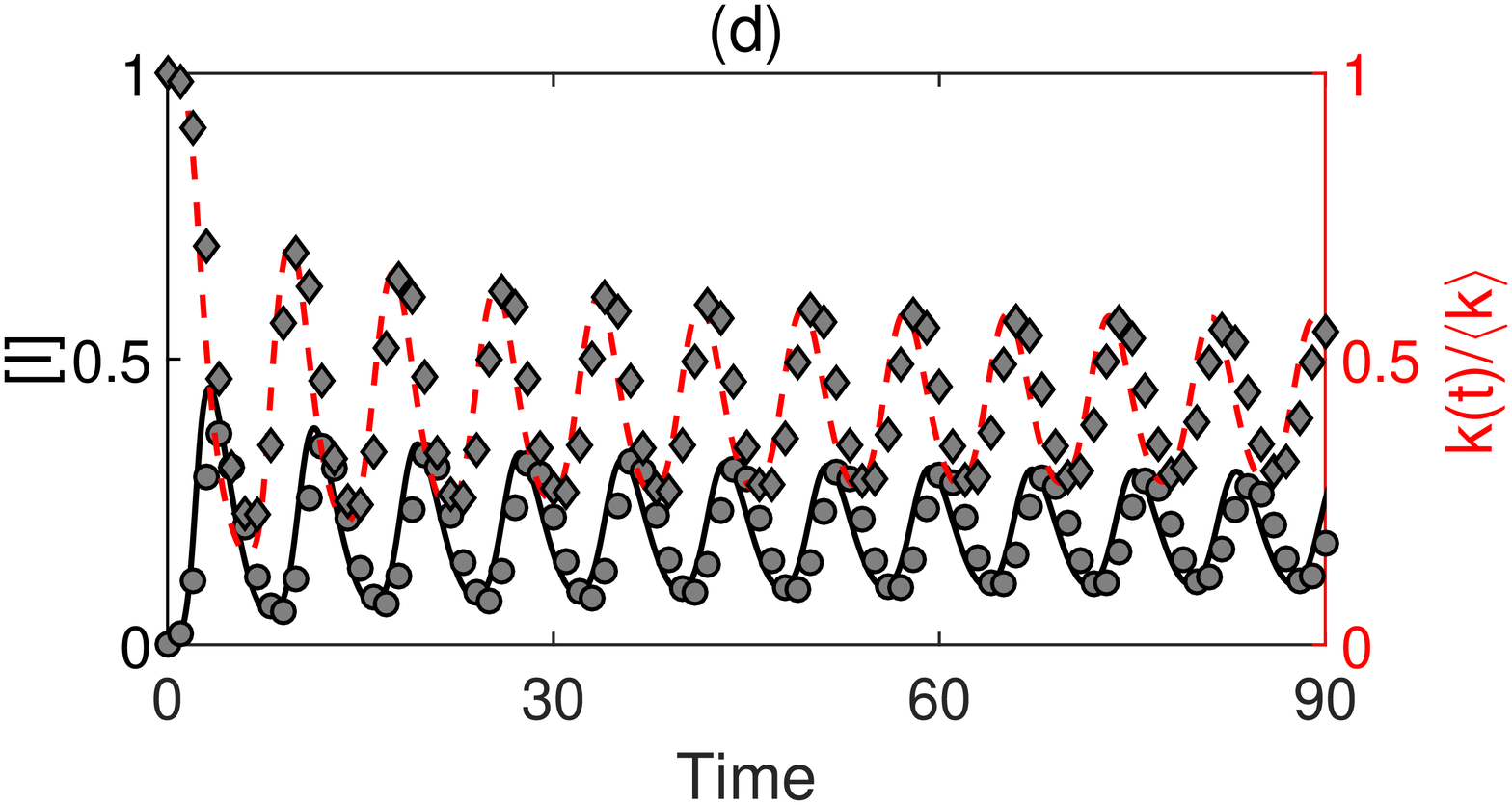}\quad
		
		\caption{(Colour online) (a) shows a two-parameter bifurcation diagram indicating different dynamical regimes in the behavior of model~\eqref{eq:sisRewire} with $\beta = 0.6$. (b) shows the \textit{endemic bubble} for model~\eqref{eq:sisRewire} for $\tau = 6$. The endemic equilibrium is stable on the grey surface and unstable on the green. Green surface is constructed using the minima and maxima of oscillations and shows the shape of the \textit{endemic bubble}. In (c) the value of the endemic equilibrium is plotted against the rewiring delay $\tau$ for $\beta = 0.6$ and $\omega = 2$. Increasing $\tau$ decreases the expected number of infected individuals at endemic equilibrium until the Hopf bifurcation point, beyond which the amplitude of oscillations grows. In (d) the average behavior from 100 numerical simulations on random networks of 1000 nodes is compared to the mean-field model~\eqref{eq:sisRewire}. The solid black line (circles) denote the prevalence of the disease in the mean-field model (simulations), and the red dashed line (diamonds) denotes the normalized mean degree calculated from~\eqref{eq:edges}. Parameter values are $\beta = 0.55$, $\omega = 1.5$, $\tau = 5.5$, $\gamma = 1$, $\ave{k}=10$. Simulations in which epidemic outbreaks died out were discarded and performed again.}
		\label{fig:group}
	\end{figure*}
	If the exponential term in \eqref{eq:Discount} is denoted by $x(t)$, the rate at which new $I-S$ edges are formed is $\omega [SI](t - \tau)(1-x(t))$. With this in mind, the mean-field model is 
	\begin{equation}
	\begin{split}
	&\dot{[S]} = -\beta[SI] + \gamma[I], \\
	&\dot{[I]} = \beta[SI] - \gamma[I], \\
	&\dot{[SS]} = 2\gamma[SI] - 2\beta\frac{[SS][SI]}{[S]} + 2\omega [SI](t - \tau)x(t), \\
	&\dot{[SI]} = -(\beta + \gamma + \omega)[SI] + \beta[SI]\left(\frac{[SS]}{[S]} - \frac{[SI]}{[S]}\right)\\
	& \hphantom{[SS] = ++} + \gamma[II] + \omega [SI](t - \tau)(1-x(t)), \\
	&\dot{[II]} = -2\gamma[II] + 2\beta\left(\frac{[SI][SI]}{[S]} + [SI]\right), \\
	&\dot{x} = -x\left\{ \gamma \ln x + \beta \left(\frac{[SI]}{[S]} - \frac{[SI](t - \tau)}{[S](t - \tau)}e^{-\gamma \tau} \right)  \right\}.
	\end{split}
	\label{eq:sisRewire}
	\end{equation}
	When $\tau = 0$, the dynamics of~\eqref{eq:sisRewire} are equivalent to the well-known model of Gross et al \cite{gross2006epidemic}.
	
	Fig.~\ref{fig:numSim} shows a comparison between the solution of the new model~\eqref{eq:sisRewire} and numerical simulation. The agreement is excellent despite the simplicity of the model and the fact that the moment closures do not reflect the changing network structure. In particular, both the solution and simulation results exhibit similar oscillatory behavior for the same parameter values. These results validate the model and allow us to analyse its behavior.
	
	Firstly, consider the basic reproductive ratio, $R_0$, defined as the expected number of secondary infections caused by a single typical infectious node in an otherwise wholly susceptible population. 
	One can find $R_0$ for the delayed rewiring model~\eqref{eq:sisRewire} via linear stability analysis near the disease-free equilibrium (DFE), $([S]^*,[I]^*,[SS]^*,[SI]^*,[II]^*, x^*)=(N,0,\langle k \rangle N,0,0, 1)$. Performing this analysis gives
	\begin{equation}
	R_0 = \frac{\beta \ave{k}}{ \gamma + \omega}.
	\label{eq:R_0Rewire}
	\end{equation}
	Note that increasing the rewiring rate decreases the epidemic threshold $R_0$, but the length of the delay, $\tau$, has no effect on the threshold. However, as we will show later, it does affect the final outcome of the epidemic. 
	
	\begin{figure*}
		\centering
\includegraphics[width=.9\textwidth]{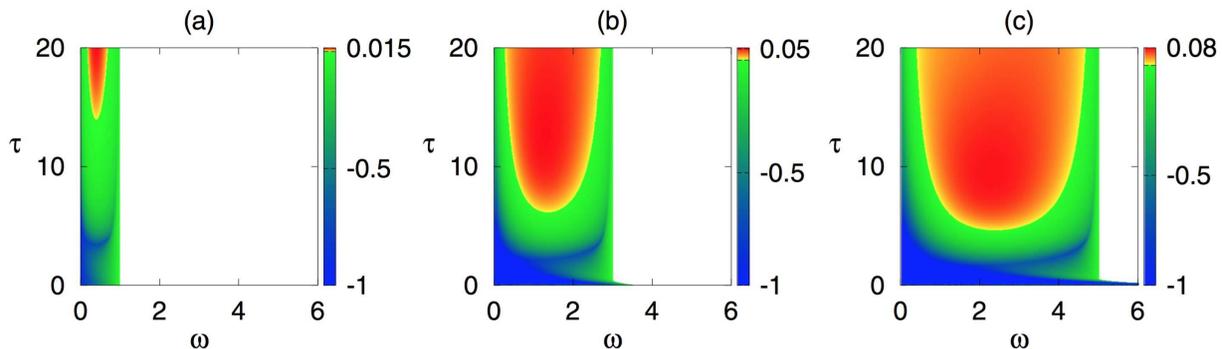}
		\caption{Real part of the maximum characteristic eigenvalue of the endemic equilibrium of~\eqref{eq:sisRewire} for $\beta = 0.2$ (a), $0.4$ (b), and $0.6$ (c). Other parameters are the same as in Fig.2 (a). The endemic equilibrium is unstable in the red/yellow region, stable in the green/blue region, and biologically infeasible in the white region.}
		\label{fig:eigs}
	\end{figure*}
	System~\eqref{eq:sisRewire} also has an endemic steady state, but its value is determined by a transcendental equation which can only be solved numerically. Using this result in the numerical linear stability analysis of~\eqref{eq:sisRewire} allows us to analyse the stability of the endemic equilibrium. As shown in Fig.~\ref{fig:group}~(a), changes to both $\tau$ and $\omega$ are capable of destabilising the endemic equilibrium. 
	Regardless of the value of $\tau$, eventually high values of the rewiring rate make the DFE stable again. For most values of $\tau$ this coincides with the point where the endemic steady state becomes biologically infeasible (less than or equal to zero), leaving the DFE as the only plausible steady state for the system. However, for sufficiently small values of $\tau$, the endemic steady state remains feasible, and there is a small region of bistability. Qualitatively, this behavior is the same for any choice of the other parameters, as long as the endemic steady state remains biologically feasible, as illustrated for different values of $\beta$ in Fig.~\ref{fig:eigs}. This figure shows that increasing the disease transmission rate results allows the endemic steady state to be feasible for a wider range of link-cutting rate $\omega$, and it also lowers the critical time delay $\tau$, at which this steady state becomes unstable.
	
	Figure~\ref{fig:group}~(b) shows the endemic equilibrium, as well as the minima and maxima of oscillations for a range of $\beta$ and $\omega$ values, with oscillations being observed in a significant part of the parameter space. One can clearly see the formation of an \textit{endemic bubble} that has been discovered earlier in other epidemic models \cite{krisztin2011bubbles,liu2015endemic}. Interestingly, both $\omega$ and $\beta$ appear to play similar roles in the formation of endemic bubble, namely they open it through a supercritical Hopf bifurcation of the endemic equilibrium and then close it through a subcritical Hopf bifurcation.
	
	Increasing the length of the delay can only induce a supercritical Hopf bifurcation, resulting in the emergence of stable oscillations, beyond which point larger values of $\tau$ only increase the amplitude of oscillations until it settles on some steady level, as shown in Fig.~\ref{fig:group} (c). One should note that the minima of oscillations get closer to zero for larger $\tau$, suggesting that for large rewiring times, there are periods of time with negligible disease prevalence, followed by major outbreaks, as illustrated in Fig.~\ref{fig:group} (d). In the limit $\tau \to \infty$, disconnected edges are never redrawn and the epidemic dies out, partially due to the network becoming sparser.
	
	For the case without time delay, Gross et al. \cite{gross2006epidemic} found bistability in a large region of the parameter space, and periodic oscillations in a much smaller region. By contrast, results shown in Fig.~\ref{fig:group} demonstrate a large region in the parameter space with oscillatory behavior. DDEs are known to often produce oscillatory dynamics, and bubbles similar to those shown in Fig.~\ref{fig:group}~(b) have been reported in other biological and epidemic models \cite{krisztin2011bubbles,liu2015endemic}.
	
	Let us now discuss the origins of oscillatory behavior in our model. The delay between disconnecting an edge and drawing a new one means that the total number of edges, and thus also the mean degree, is not constant. Whenever a susceptible node chooses to rewire, the total number of edges in the network decreases by two (since all edges are bidirectional) until time $\tau$ passes, and the edge is redrawn. The mean degree $k(t)$ at any time $t$ can be calculated directly from this argument as follows,
	\begin{equation}
	k(t) = \ave{k} - 2\omega \int_{t - \tau}^t [SI](u) \, du.
	\label{eq:edges}
	\end{equation}
	Figure~\ref{fig:group}~(d) shows that oscillations are driven by the dynamics of $k(t)$. During the early stages of an outbreak with a high rewiring rate $k(t)$ falls rapidly, as susceptible nodes cut links in response to the propagation of the disease. If the value of $\tau$ is large enough, then after a certain time the number of edges in the network is small enough to effectively starve the disease of transmission routes, and prevalence falls. These edges are then redrawn at the same rate as they were cut $\tau$ time ago, and $k(t)$ grows, which allows the disease to spread again. Figure~\ref{fig:group}~(d) illustrates this behavior both in simulation and in the mean-field model~\eqref{eq:sisRewire}, showing how after the initial outbreak each new wave of infection is preceded by the recovery of network connectivity.
	
	\begin{figure*}[htp]
		\includegraphics[width=\textwidth]{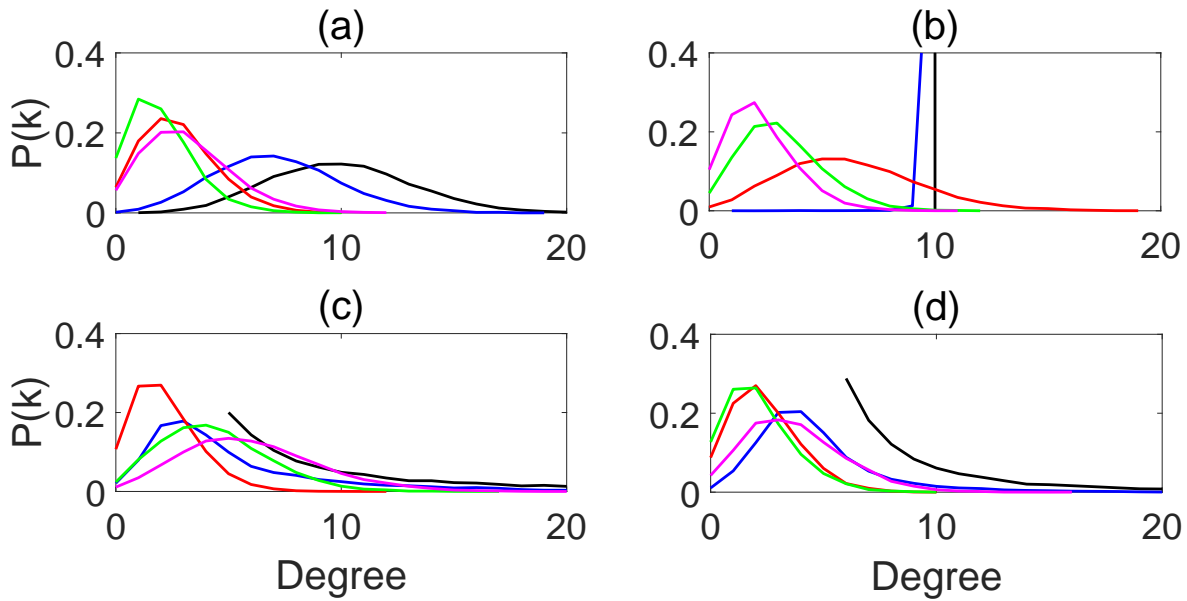}
		\caption{(Color online) Snapshots of the degree distribution for networks of $10^4$ nodes. In each plot the solid black solid is the initial degree distribution, the blue line is for the early growth phase, red shows an early peak, and green and magenta later snapshots. Disease parameters are $\beta = 0.6$, $\gamma = 1$, $\omega = 1.4$, $\tau = 6$. (a) Erd{\H o}s-R{\'e}nyi network with $\ave{k} = 10$. (b) Homogeneous network with $k = 10$ for all nodes. (c), (d) Truncated scale-free networks with the scaling exponents $\alpha = 2$ and $\alpha = 3$, respectively.}
		\label{fig:snapshots}
	\end{figure*}
	The effect of oscillatory interactions between network connectivity and the propagating disease may be more pronounced in network simulations. Gross et al. \cite{gross2006epidemic} found that adaptive rewiring without delay can lead to the formation of highly connected clusters of susceptible nodes that are vulnerable to disease once any one node becomes infected. Since the model~\eqref{eq:sisRewire} does not account for changes in network structure, i.e. the closure is the same for all times and it does not depend on the average degree or degree distribution, this can potentially explain the small discrepancy between the solution of the deterministic and simulation models observed in Fig.~\ref{fig:numSim}.
	
	To get a better understanding of the interplay between network topology and dynamics, it is worth looking at how delayed rewiring alters degree distribution. Time snapshots of several large networks in Fig.~\ref{fig:snapshots} show the evolution of the degree distribution at various key points of an epidemic in an oscillatory regime. The initial network topology (black lines) is quickly reorganised to a peaked distribution. The oscillations in prevalence cause slight but repeated changes in the degree distribution. Unsurprisingly, when prevalence is at or near its peak, nodes with a lower degree are more common. When the prevalence falls, the distribution curves shift to the right, and the shape of the distribution flattens slightly. When the endemic steady state is stable, the degree distribution stabilises to a peaked distribution between the two extremes of the oscillatory regime. A very important observation is that irrespective of the initial network topology, due to rewiring different networks eventually settle on a very similar skewed degree distribution. This implies that earlier conclusions derived for the specific closure (\ref{clos}) appropriate for Erd{\H o}s-R{\'e}nyi graphs are actually applicable to modeling long-term dynamics of different types of networks, for which the influence of the initial topology is low since significant amount of rewiring has already taken place.
	
	The particular strength of this model lies in its ability to exhibit rich behavior from a simple system of DDEs. Time delay captures the fact that finding alternative contacts takes time, and also during an epidemic many people try to temporarily reduce the number of their contacts. Such behavior can be modelled using this delayed rewiring process. Previous work separated the processes of edge destruction and creation, and with edge creation occurring at a fixed rate, the number of edges in the network was bounded only by the network size \cite{althouse2014epidemic, huepe2011adaptive}. In the new model presented above, edge creation is reduced to replenishing global network connectivity towards its original level. Therefore, this model is fundamentally different to those earlier models, even when parameters are matched.
	
	During the initial growth phase it is the rate at which potential transmission is avoided by cutting a link, not the delay before drawing a new edge, that determines whether a major outbreak will occur. Although the delay does not affect the basic reproductive ratio $R_0$, it does impact the outcome of the epidemic (see Fig.~\ref{fig:group}~c). The result of introducing the delay is that oscillations occur in a large region of the parameter space. This happens due to the interplay between the spread of the disease and the behavioral changes in response to the epidemic. When the length of the delay is significant, the network becomes more sparse, healthy individuals are at lower risk of infection, and over time the prevalence falls. When the new edges are then formed, the disease is once again able to spread, and the cycle repeats.
	
	Understanding the nature and cause of oscillations may provide opportunities to eradicate the disease. For example, if public awareness campaigns can lead to an increase in the length of the delay, the prevalence of the disease will naturally fall close to zero, at which time a relatively minor intervention, such as quarantining those who remain infected, may be enough to eradicate the disease from the population entirely.
	
	Currently, the model assumes that only susceptible nodes rewire. However, in reality, infected nodes are also likely to change their behavior. Risau-Gusman and Zanette \cite{risau2009contact} considered a model of rewiring where infected nodes rewire with a given probability. It would be of great value to examine a similar situation under delayed rewiring, with time delay representing the time for which infected nodes partially isolate themselves before rewiring, in accordance with advice given by public health authorities. This would alter the nature of the variable $x(t)$ in the model. For example, if only infected nodes rewire, $x(t) \approx e^{-\gamma \tau}$. Preliminary tests of this rewiring scheme show behavior similar to the present model.
	
	Numerical simulations have shown that a similar oscillatory behavior is observed for other initial network topologies, including scale-free networks. Furthermore, since rewiring nodes choose their new neighbours uniformly at random from all available susceptible nodes, the initial network topology itself is transient, as shown in Fig.~\ref{fig:snapshots}, and, as a result, over time our model becomes more relevant. Future work will look at how the degree distribution and oscillations are affected in the case when the network links are rewired not randomly but according to a preferential attachment or some fitness-based rule. This could result in some interesting new dynamics due to the competition between the increased probability of highly-connected nodes receiving new links, and the increased probability of infection.
	
	\begin{acknowledgments}
		The authors are grateful to anonymous reviewers for their helpful comments and suggestions. N. Sherborne acknowledges funding for his PhD studies from the EPSRC, EP/M506667/1, and the University of Sussex.
	\end{acknowledgments}
	
%

\end{document}